\documentclass[useAMS,usenatbib]{mn2e}
\usepackage{dblfloatfix}

\usepackage{mathtext,amssymb}
\usepackage{epsfig}
\usepackage{graphics}
\usepackage{url}

\usepackage{times}

\renewcommand{\div}{\ensuremath{\mbox{\,--\,}}}

\newcommand{\Msun}{\ensuremath{\mathrm{M_\odot}}}
\newcommand{\Msunyr}{\ensuremath{\mathrm{M_\odot\,yr^{-1}}}}

\newcommand{\kms}{\ensuremath{\mathrm{km\,s^{-1}}}}

\newcommand{\ergf}{\ensuremath{\mathrm{erg\,cm^{-2}\,s^{-1}\,Hz^{-1}}}}
\newcommand{\Gpc}{\ensuremath{\mathrm{Gpc}}}

\newcommand{\pardir}[2]{\frac{\partial}{\partial #1} \left( #2\right)}

\newcommand{\der}[2]{\frac{d}{d#1} \left( #2\right)}

\newcommand{\sbs}{SBS~J1520+530}
\newcommand{\einc}{QSO~J2237+0305}
\newcommand{\pg}{PG~1115+080}
\newcommand{\rxn}{RXJ~0911+0551}
\newcommand{\rxj}{RXJ~1131-1231}
\newcommand{\sdss}{SDSS~0924+0219}

\newcommand{\mg}{MG~J0414+0534}

\newcommand{\wf}{WF1~J2033-4723}
\newcommand{\grs}{GRS~J1915+105}

\newcommand{\mdot}{\ensuremath{\dot{m}}}

\newcommand{\qone}{Q~J0158-4325}
\newcommand{\heone}{HE~J0435-1223}
\newcommand{\fb}{FB~J0951+2635}
\newcommand{\sdten}{SDSS~J1004+4112}
\newcommand{\heel}{HE~J1104-1805}
\newcommand{\sdel}{SDSS~J1138+0314}

\newcommand{\AAA}{\ensuremath{\rm \AA}}

\newcommand{\atan}{{\rm atan\,}}

\title[Microlensing Evidence for Super-Eddington Accretion]{Microlensing Evidence for
  Super-Eddington Disc Accretion in Quasars}
\author[Abolmasov \& Shakura]{P.\ Abolmasov\thanks{E-mail:
pavel.abolmasov@gmail.com} and N.\ I.\ Shakura, \\
Sternberg Astronomical Institute, Moscow State University,
  Universitetsky pr., 13, Moscow, 119992, Russia\\
}
\begin{document}

\date{Accepted ---. Received ---; in
  original form --- }

\pagerange{\pageref{firstpage}--\pageref{lastpage}} \pubyear{2009}

\maketitle

\label{firstpage}

\begin{abstract}
Microlensing by the stellar population of lensing galaxies provides an
important opportunity to spatially resolve the accretion disc structure in
strongly lensed quasars. Disc sizes estimated this way are on average larger
than the predictions of the standard Shakura-Sunyaev accretion disk
model. Analysing the
observational data on microlensing variability allows to suggest that some
fraction of lensed quasars (primarily, smaller-mass objects) are accreting in
super-Eddington regime. Super-Eddington accretion leads to 
formation of an optically-thick envelope scattering the radiation formed in
the disc. This makes the apparent disc size larger and practically independent
of wavelength. In the framework of our model, it is possible to make
self-consistent estimates of mass accretion rates and black hole masses for
the cases when both amplification-corrected fluxes and radii are available. 
\end{abstract}


\section{Introduction}

Since the work of \citet{lyndenbell69}, disc accretion onto supermassive
black holes is a commonly accepted interpretation for the activity of quasars,
both radio-loud and radio-quiet (sometimes distinguished as quasi-stellar
objects, QSO). Among all the active galactic nuclei, quasars are
distinguished by higher luminosities (exceeding that of host galaxies)
that is most likely connected to higher accretion rates. 

Spectral energy distributions in optical and UV are reasonably consistent
\citep{elvis_seds} with
the predictions of the standard thin accretion disc model introduced in
the seminal works of \citet{shakura72,SS73, LP74,NT73}. However, the
predicted angular sizes of quasar accretion discs are too small
(microarcseconds and less) to be resolved directly.
For today, quasars
remain essentially point-like (``quasi-stellar'') objects resolved only
indirectly, in particular by microlensing effects.

As it was shown by \citet{agolkrolik99}, microlensing by the stellar population of lensing
galaxies is sensitive to the size of the emitting region. 
Here, we adopt the statement of \citet{size} that the basic quantity
that microlensing amplification maps and curves are sensitive to is \emph{half-light
radius}. Half-light radius $R_{1/2}$ is defined as the radius inside which half of the
observed flux is emitted at a given wavelength. 


Numerous studies aimed on probing the spatial properties of quasar accretion
discs with help of
microlensing. While most early works (see \citet{wambsganss} for review)
reported reasonable agreement
between the observational data and the standard accretion disc theory, several
important results are at odds with the theoretical predictions. Studying
microlensing amplification statistics, \citet{pooley07} find best-fit disc
sizes more than one order of magnitude larger than the 
theoretical predictions based on photometrical data.
Partially, this may be attributed to the mass
estimates used in this study (see discussion in \citet{paper1}). In
\citet{morgan10}, inconsistency is somewhat smaller (about a factor of 3) but
still significant. Accretion discs seem too large for their apparent
luminosities or too faint for their sizes in the
UV/optical range ($\sim 2000\div 4000\AAA$). Many papers (such as
\citet{jimenez}, \citet{pooley07} and \citet{morgan10}) interpret this
inconsistency as an evidence for insufficiency
of the standard accretion disc model, but no universal solution was proposed
so far to account for the size discrepancy. There are indications for possibly
higher black hole masses in some objects \citep{morgan10,paper1}, but no
changes in masses, accretion rates and efficiencies can explain the
observed sizes and fluxes simultaneously. 

One of the important issues in quasar microlensing studies is
whether the disc radial scale $R_S$ \citep{morgan10} 
dependence on wavelength is consistent with the power law $R_S
\propto \lambda^{4/3}$ predicted by the standard accretion disc theory. Several important
disc models predict power law dependences $R(\lambda) \propto \lambda^\zeta$
with different exponents. Below we will refer to $\zeta$ as ``structure parameter''.  
While for \einc, classical $\zeta=4/3$ works fairly good
\citep{eigenbrodII,anguita08}, other objects such as
\sdss, for instance, clearly require smaller $\zeta$. \citet{floyd09}
propose angular momentum inflow
at the inner edge of the disc in \sdss\ \citep{agolkrolik} as an explanation for the
apparently very steep temperature law in the disc. This model implies $R_S
\propto \lambda^{8/7}$ marginally consistent with the observational data. 

In the recent work by \citet{blackburne}, several other objects were shown to
have much shallower $R(\lambda)$ dependences, some consistent with $R_S =
const$ for a broad range of comoving wavelengths, $0.1\mu \lesssim \lambda
\lesssim 1\mu$. The only object having conventional thin-disc scaling is
\mg\ that has the highest mass among the sample of 12 objects considered by
\citet{blackburne}. All the smaller-mass ($M\lesssim 10^9\Msun$) black holes
are characterised by $\zeta \sim 0\div 0.5$. 

{
Microlensing effects in the X-ray range are more profound than in the optical \citep{pooley07}.
Independently of the disc structure in the optical and UV ranges, X-ray
properties are more or less similar for all the objects where microlensing
effects were studied in the X-ray range \citep{einc_xrays,
  dai10,morgan12}. Evidently, X-ray emission comes from somewhere inside the
inner $\sim 10\times GM/c^2$ \citep{chen12}, and the exact mechanisms driving
the formation of the X-ray continuum and lines are yet to be revealed. 
}

Small structure parameters originate not only for very steep temperature
slopes in multi-blackbody models. For instance, $\zeta =0$ is naturally
reproduced if the brightness distribution does not depend on
wavelength. This may be achieved if the accretion disc is surrounded by an
envelope optically thick to Thomson scattering. Without affecting its spectral
properties, scattering changes the spatial brightness distribution of the
disc radiation.
In general, accretion disc will increase its apparent radius and lose its
intrinsic radius dependence on wavelength. 
A possible origin for such a scattering envelope is super-Eddington accretion that
leads to formation of a Thomson-thick wind \citep{SS73}. 
Since there is observational evidence that some quasars accrete in supercritical regime,
especially at larger redshifts \citep{collin}, we consider this scenario
quite plausible. 

In the following section \ref{sec:sph}, we describe a simple 
scattering envelope model that we use to account for the spatial
properties of microlensed quasars. It will be shown that such an envelope
may result from super-Eddington accretion by a supermassive black hole. 
In section \ref{sec:obs} and
\ref{sec:res}, we describe the observational data and interpret them in the
framework of the scattering envelope model. In section \ref{sec:balq}, we
consider the possible connection between the putative class of supercritical
quasars, broad absorption line (BAL) quasars, and make conclusions in
section \ref{sec:conc}. 

\section{Spherical envelope model}\label{sec:sph}

{
Though the issue of super-Eddington accretion is complicated, and different
effects like photon trapping should be taken into account, the
simple picture of supercritical accretion introduced by \citet{SS73} is
sufficient for our needs. 
This picture is supported by comprehensive numerical simulations
\citep{ohsuga,ohsuga11}.
}

\subsection{Transition to supercritical regime}\label{sec:tran}

Radiation pressure is the principal feedback source for disc
accretion. While for a spherically-symmetric source, both radiation pressure
force and gravity scale $\propto R^{-2}$ with distance that leads to the
universal Eddington luminosity limit, accretion disc geometry makes the
situation more complicated because the source is no longer isotropic, and the
forces are no longer collinear. For a thin disc, vertical gravity component
grows with the vertical coordinate $z$ while the flux generated in the disc
does not significantly depend on $z$ high enough in the atmosphere. Thin
accretion disc supported by radiation pressure will have thickness
determined by the equilibrium of the vertical components of the two forces. 

$$
\frac{\varkappa}{c}F = \frac{\varkappa}{c} \frac{3}{8\pi} \frac{GM\dot{M}}{R^3} \left(
1-\sqrt{\frac{R_{in}}{R}} \right) =  \frac{GM}{R^3} H
$$

Here, $R$ is the radial coordinate, 
$R_{in}$ is the inner disc edge that for a black hole disc it is 
instructive to identify with the innermost stable orbit radius, $\varkappa
\sim 0.4\rm cm^2 g^{-1}$ is electron scattering opacity, $c$ is the speed of
light, $G$ is gravitational constant, $M$ and $\dot{M}$ are the black hole
mass and accretion rate. 
$H$ is accretion disc half-thickness that may be subsequently expressed
as:

$$
H=\frac{3}{8\pi} \frac{\varkappa \dot{M}}{c} \left(1 -
\sqrt{\frac{R_{in}}{R}}\right) = \frac{3}{2} \frac{GM}{c^2} \mdot \left( 1 - \sqrt{\frac{R_{in}}{R}}\right)
$$

Here, we normalised the mass accretion rate as $\dot{M} = \dot{M}^* \mdot$, 

\begin{equation}\label{E:Mcr}
\dot{M}^* = \frac{L_{Edd}}{c^2} = \frac{4\pi GM}{c\varkappa},
\end{equation}

where $L_{Edd} = 4\pi GMc / \varkappa$ is the Eddington luminosity. 
When the thickness of the radiation-supported disc becomes comparable to its
radius, thin-disc approximation breaks down and the balance of forces is
inevitably shifted: gravity scales as $(z^2+R^2)^{-1}$, while the disc
radiation decreases more slowly because at these distances accretion disc is
still a strongly extended radiation source. Hence we assume the disc
super-Eddington if its equilibrium half-thickness is $H\geq R$. 
This condition defines the \emph{spherisation radius} in a non-relativistic
regime but with a
correction term that may play non-negligible role near the critical accretion
rate. Let us introduce dimensionless radius $r=R/R_{in}$,
where $R_{in} = x_{in}\times GM/c^2$ is the inner radius of the
disc. Dimensionless inner radius $x_{in}$ varies between 1 (extreme
Kerr case, corotating disc) and 9 (extreme Kerr, counter-rotation). 
Existence of a supercritical region in the disc requires existence of a root
$r>1$ of the following equation:

\begin{equation}\label{E:rsph}
\frac{r}{1-1/\sqrt{r}} = \frac{3}{2} \frac{\mdot}{x_{in}}
\end{equation}

This equation may be reduced to a cubic equation for $\sqrt{r}$ 
having either two or no positive real roots, depending on the right-hand side. 
The minimum of the left
hand side is always at $r_{cr} =9/4$. This implies the critical accretion rate
of $\mdot_{cr} = 4.5x_{in}$. The luminosity of a non-relativistic disc with
this value of mass accretion rate is $L/L_{Edd} =  \mdot_{cr} \eta \simeq
9/4$, where $\eta = L/\dot{M}c^2$ is accretion efficiency. 
This value may be thought of as the gain that disc geometry provides for the
Eddington limit. Without the correction term, the critical accretion rate should be
about an order of magnitude lower, $\mdot_{cr,0} = 2x_{in}/3$. This
justifies the attention we pay to the correction term. Still, the real
transition to supercritical accretion may be much more complicated and
influenced by relativistic effects and the exact physics of wind
acceleration. 

Spherisation radius for $\mdot > \mdot_{cr}$ may be found in a way that takes
into account the correction term:

\begin{equation}\label{E:rsph:sol}
R_{sph} = \frac{3}{2} \frac{GM}{c^2} \mdot \times \psi^2(\mdot/x_{in})
\end{equation}

Here, $\psi$ is the correction multiplier that accounts for the influence of
the inner radius in the disc. It may be found as the largest root of the cubic equation
for $\sqrt{r}$ that follows from (\ref{E:rsph}). When accretion is
supercritical ($\mdot > \mdot_{cr} = 4.5 x_{in} $), the cubic equation has three roots
and its solution may be expressed using trigonometric functions
\citep{cubic}. Solving the cubic equation yield the following expression for
correction multiplier:

\begin{equation}\label{E:psi}
\psi(x) = \frac{2}{\sqrt{3}} \cos\left( \frac{1}{3} \arccos\left(-\frac{3}{\sqrt{2x}}\right) \right) 
\end{equation}

For $\mdot\gg 1$,  $\psi(\mdot/x_{in})$ rapidly approaches $1$, hence:

$$
R_{sph} \simeq \frac{3}{2} \mdot \times  \frac{GM}{c^2},
$$

that coincides with the classical definition of spherisation radius \citep{SS73}.




Formation of disc winds in supercritical accretion regime is supported by
numerical simulations \citep{ohsuga,okuda}.
If the radiation flux from the disc exceeds the local Eddington value, some
part of its energy is converted to kinetic energy of the outflow.
Terminal wind
velocity $v_w$ may be estimated through Bernoulli integral:

$$
\frac{v_w^2}{2}
\simeq \frac{GM}{R} \times \left(\frac{ L
}{L_{Edd}} - 1\right) 
$$

For example, setting the luminosity to the ``disc Eddington limit'' $L=
\frac{9}{4} L_{Edd}$  implies $v_w = \sqrt{2.5 GM/R_w}$ where  $R_w$ is some
effective radius where the outflow is formed. 
Since most of the matter is ejected from $R\sim R_{sph}$, we
assume the terminal velocity of the wind is proportional to the escape
velocity at the spherisation radius:

\begin{equation}\label{E:vw}
v_w=\beta \sqrt{2GM/R_{sph}},
\end{equation}

were $\beta\sim 1$ is some additional dimensionless multiplier introduced to
account for the
uncertain details of wind acceleration and formation of the outflow.

\subsection{Spherical envelope radius}\label{sec:rsph}

Let us assume that the envelope is composed of a fully ionised
spherically-symmetric wind expanding at a constant velocity. 
Continuity equation allows to connect electron density with the mass loss rate
$\dot{M}_w = f_w \mdot \dot{M}^*$ and outflow velocity $v_w$ calculated
according to equation (\ref{E:vw}). 

$$
\rho = \frac{\dot{M}_w}{4\pi R^2 v_w}
$$

Envelope size is defined by the radius where the radial optical
depth toward the observer is unity. 

$$
\tau(R) = \int_0^{+\infty} \varkappa \rho(R) dR = \frac{f_w \varkappa
  \dot{M}}{4\pi c v_w R}
$$

$$
R_1 = R(\tau=1) = \frac{f_w \varkappa \dot{M}}{4 \pi v_w } 
$$

\begin{equation}\label{E:r1}
\frac{R_1 c^2}{GM} =  \sqrt{\frac{3}{8}} \frac{f_w}{\beta } \mdot^{3/2} \psi(\mdot/x_{in})
\end{equation}

For a moderately super-Eddington accretor with $f_w \sim 1$, the size of the
envelope becomes comparable and may even exceed the size of the accretion disc
in the ultraviolet range. Such a scattering envelope has a radius practically
independent of wavelength while the spectral properties of the scattered
radiation remain more or less
unchanged. The envelope is actually a pseudo-photosphere: it expands
supersonically at a large, possibly mildly relativistic velocity of $v \sim
\mdot^{-1/2} c$. 

\subsection{Apparent intensity distribution}

Half-light radius is defined by the general relation that may be used for any
radially-symmetric intensity distribution:

\begin{equation}\label{E:halflight}
\frac{\int_{R_{in}}^{R_{1/2}} I(R) R dR}{\int_{R_{in}}^{+\infty} I(R) R dR}=\frac{1}{2}
\end{equation}

For a standard thin non-relativistic accretion disc, monochromatic intensity
scales with radius as:

$$
I(R) \propto \frac{1}{\exp\left( \left(R/R_S\right)^{3/4}\left(1-\sqrt{\frac{R_{in}}{R}}\right)\right)-1},
$$

where $R_S$ is the disc radial scale defined by condition $h\nu_{em} =
kT(R_S)$ without the correction factor. If $R_{in} \ll R_{1/2}$, to a high
accuracy:

\begin{equation}\label{E:rd}
\begin{array}{l}
R_{1/2}^{(disc)} 
\simeq 2.44 R_S \simeq \\
\qquad{} \simeq 2.44 \left(\frac{45 \lambda_{em}^4 GM\dot{M}}{16\pi^6 h c^2}
\right)^{1/3} = \\
\qquad{} = 2.44 \left(\frac{45 c^3 \lambda_{em}^4 \mdot }{4\pi^5 h \varkappa
  GM}\right)^{1/3} \frac{GM}{c^2}\\
\end{array}
\end{equation}

Here, $\lambda_{em}$ is the comoving (quasar reference frame) 
wavelength of the observed radiation ($\nu_{em} = c/\lambda_{em}$ is
corresponding frequency; observed wavelengths and frequencies are denoted as
$\lambda_{obs}$ and $\nu_{obs}$), $h$ is Planck constant. 

For a spherical envelope, brightness is nearly uniform in the centre and
declines as a power law at large radii. We have taken the extended
scattering  
atmosphere model described in Appendix~\ref{sec:app} and calculated the intensity at
infinity for different shooting parameter values coming to the overall
conclusion that the half-light radius in this model is proportional to the
photosphere radius and $R_{1/2}^{(envelope)} \simeq 1.06 R_1$. 

\subsection{Disc radiation}

Standard accretion disc temperature law may be written (neglecting the
correction term important for the inner parts of the disc) as:

$$
T(R) = \left( \frac{3}{2} \frac{G^2M^2}{\sigma \varkappa c} R^{-3} \mdot \right)^{1/4},
$$

where $\sigma$ is Stephan-Boltzmann constant. 
Monochromatic flux is found as an integral over the picture plane:

$$
F_\nu = \int I_\nu d\Omega = \frac{2\pi \cos i}{D^2\times (1+z)^3} \int
I_\nu(R) R dR
$$

We use angular size distance $D=D(z)$, and $i$ is disc inclination. 
Assuming the radiation generated in the disc has locally a blackbody spectrum
leads to the
following estimate for monochromatic flux valid far away from the high- and
low energy cut-offs connected to the inner and the outer disc edges:

\begin{equation}\label{E:fnu}
\begin{array}{l}
F_\nu =  8\pi \left(\frac{2}{3}\right)^{1/3} \Gamma(8/3)\zeta(8/3)
\frac{k^{8/3}\nu_{obs}^{1/3}}{c^{8/3}h^{5/3}\varkappa^{2/3}\sigma^{2/3}}\times \\
\qquad{} \times
(GM)^{4/3} \mdot^{2/3} \cos i \times \frac{1}{D^2\times
  (1+z)^{8/3}} \\
\end{array}
\end{equation}

Here, $\Gamma$ and $\zeta$ are Gamma-function and Riemann zeta-function (not
to be confused with the structure parameter $\zeta$ that is used without any
argument). 
The above formula may be used to estimate the mass of the supermassive black
hole (SMBH) using one
observable quantity (flux) and one unknown parameter \mdot. 

\begin{equation}\label{E:fnumass}
\begin{array}{l}
M = \left(\frac{3}{2} \right)^{1/4} \times (8\pi \Gamma(8/3)
\zeta(8/3))^{-3/4} \times \frac{c^2 h^{5/4} \varkappa^{1/2}
  \sigma_B^{1/2}}{Gk^2 \nu_{obs}^{1/4}} \times \\
\qquad{} \times F_\nu^{3/4} \mdot^{-1/2} \cos^{-3/4} i D^{3/2}(z) \times
(1+z)^2 
\end{array}
\end{equation}

Spherical envelope scrambles the radiation generated in the disc and makes it
roughly isotropic. Effective inclination cosine in this case is $\cos i_{eff}
= 1/2$ since the initially anisotropic flux $F_\nu \propto \cos i$ is
re-distributed isotropically with the total luminosity conserved.

\subsection{Mass and accretion rate estimates}\label{sec:mmdot}

Microlensing studies are unique for distant quasars since they allow to
estimate the size of the
emitting region in continuum independently of the observed flux. 
In the case of a standard accretion disc both
observables, $F_\nu$ and $R_{1/2}$, may be used to estimate only one
combination of black hole mass and accretion rate, $M^2\mdot$ (see also
discussion in
\citet{paper1}). Existence of a scattering envelope allows to break this
degeneracy and make self-consistent estimates of both principal parameters
($M$ and $\mdot$). 


Solving the system of two equations (\ref{E:fnumass}) and (\ref{E:r1}) for $M$
and \mdot\ allows to estimate both black hole mass and dimensionless accretion
rate. We also set $\cos i = \cos i_{eff}=1/2$:

\begin{equation}\label{E:mdotiter}
\begin{array}{l}
\mdot = \frac{\sqrt{2}}{3^{3/4}} \left(8\pi \Gamma(8/3)\zeta(8/3)
\right)^{3/4} \frac{k^2\nu_{obs}^{1/4}}{h^{5/4} \varkappa^{1/2} \sigma_B^{1/2}
}\times \\
\qquad{} \times F_\nu^{-3/4} D^{-3/2}(z) (1+z)^{-2}  \times
\frac{\beta}{f_w} \frac{R_1}{\psi(\mdot/x_{in})} \simeq \\
\qquad{} \simeq 169 \left(\frac{\lambda_{obs}}{0.79\mu}\right)^{-1/4} \times 10^{0.3(I-19)} \times 
\left(\frac{D}{1\Gpc}\right)^{-3/2} \times \\
\qquad{} \times (1+z)^{-2} \times \frac{\beta}{f_w} \frac{R_1}{10^{15}\rm cm} \frac{1}{\psi(\mdot/x_{in})} \\
\end{array}
\end{equation}

Here, we expressed the observed flux through the magnitude $I$ in the {\it
  HST} F814W band ($\lambda_{obs} \simeq 0.79\mu$) adopting the zero-point
flux equal to $F_\nu = 2.475\times
10^{-20}\ergf$ \citep{holtzman} since we use the amplification-corrected
magnitudes from \citet{morgan10} obtained in this photometrical band at the
{\it HST}. 

Since the left-hand side of (\ref{E:mdotiter}) dependence on the mass accretion
rate is much stronger, simple forward iteration works good.  Once
\mdot\ is found, the black hole mass may be estimated following
(\ref{E:fnumass}) as:

$$
M \simeq 4.6\times 10^7 \times 10^{-0.3 (I-19)} \left(\frac{D(z)}{1\Gpc}
\right)^{3/2} (1+z)^2 \mdot^{-1/2} \Msun
$$

\subsection{Disc and envelope sizes}

Depending on the wavelength range, a supercritical disc 
may be either observed directly (if its size is larger than the photosphere of
the wind) or covered by the photosphere of the supercritical wind. 
Equality of the half-light radii set by the
two radial scales leads to the following condition for observational importance
of the envelope:

$$
R_{1/2}^{(disc)} = R_{1/2}^{(envelope)}
$$

$$
2.44 R_S = 1.06 R_1
$$

After substituting equations (\ref{E:rd}) and (\ref{E:r1}), one gets the
following mass limit:

\begin{equation}\label{E:mthresh}
\begin{array}{l}
 M_{lim} = \left( \frac{2.44}{1.06}\right)^3 \left(\frac{4}{3}\right)^{3/2}
 \frac{45}{4\pi^5}  \left(\frac{\beta}{f_w}\right)^3 \times \\
\qquad{} \times \frac{c^3}{
  \varkappa h G} \lambda_{em}^4 \mdot^{-7/2} \psi^{-3}(\mdot/x_{in})
\simeq \\
\qquad{} \simeq 3.8 \times 10^{10} \left(\frac{\lambda_{em}}{0.25\mu}\right)^4
\left(\frac{\mdot}{10}\right)^{-7/2} \psi^{-3}(\mdot/x_{in}) \Msun
\end{array}
\end{equation}

For higher masses (at a given wavelength, for fixed \mdot), the size of the
envelope is smaller than the half-light radius of the disc, and the appearance
of the quasar will be close to the thin disc case. The limit is shown in
figures \ref{fig:mmdot} and \ref{fig:mmdotphys} with solid lines. This limit
evidently depends on wavelength. For the
sample of \citet{morgan10}, comoving frame wavelength changes in the range
$0.2\div0.5\mu$ that corresponds to about a factor of 2 upward shift
of the limit in figure \ref{fig:mmdot}.

\section{Multi-wavelength data}\label{sec:obs}


\subsection{Disc radii and amplification-corrected fluxes}\label{sec:obs:rad}

We make use of the amplification-corrected fluxes and microlensing-based
radii collected and published by \citet{morgan10}. The sample of
\citet{morgan10} overlaps with this of \citet{blackburne}. In the original
work, disc radii are given in terms of accretion disc radial scale $R_S$. 
Since we propose that at least for some
objects, emitting region has different nature and intensity distribution, 
we recalculate these disc radii
into model-independent half-light radii. Fitted with a standard-disc model
with radial scale $R_S$ as defined above, emitting region
may be characterised by the half-light radius of $R_{1/2}
\simeq 2.44 R_S$.  The scattering 
envelope model we use has one characteristic radius $R_1$ where optical
depth equals unity. For this model, $R_{1/2} \simeq 1.06 R_1$ (see
Appendix~\ref{sec:app}). Half-light radii are given in table
\ref{tab:pro}. For all the objects studied by \citet{morgan10} and
\citet{blackburne}, the two estimates are consistent within the
uncertainties. Since the uncertainties are very high, it is difficult to set
any constraints upon the possible variations of the half-light radii. 

\begin{table*}\centering

\caption{Properties of the microlensed quasars from the sample of 
  \citet{morgan10}. Black hole mass $M_S$ and accretion rate $\mdot$ are
  calculated for the case of supercritical accretion for $a=0.9$. 
}
\label{tab:pro}

\bigskip
\begin{tabular}{lcccccc}

Object & $M_{vir}, 10^9\Msun$  &  $I_{corr}, mag$ & $R_{1/2}$, $10^{15}\rm cm$ 
 & $R_{1/2}$, $10^{15}\rm cm$ & $M_S, 10^9\Msun$ & $\mdot$ \\
       &                         &                  &  \citep{morgan10} &
       \citep{blackburne} &   & \\             
\hline
\heone & 0.50  & 20.76$\pm$0.25 & $12.23_{- 26.44}^{+ 9.79}$ & $9_{-5}^{+12}$ &$0.028^{+0.03}_{-0.013}$ & $70^{+90}_{-40}$ \\
\sdss & 0.11& 21.24$\pm$0.25 & $ 2.44_{- 2.43}^{+ 1.47}$ & $ 4.6_{-2.4}^{+5} $ & $0.026^{+0.017}_{-0.009}$ & $32^{+20}_{-15}$ \\
\fb & 0.89& 17.16$\pm$0.11 & $ 30.72_{- 46.44}^{+ 18.49}$ & -- &$0.33^{+0.15}_{-0.11}$ & $31^{+30}_{-14}$ \\
\sdten  & 0.39& 20.97$\pm$0.44 & $ 1.94_{- 1.93}^{+ 0.97}$ & -- &$0.044^{+0.02}_{-0.015}$ & $22^{+15}_{-8}$ \\
\heel & 2.37 & 18.17$\pm$0.31 & $ 19.38_{- 11.34}^{+ 9.67}$ & -- & $0.42^{+0.25}_{-0.14}$ & $23^{+12}_{-9}$ \\
\pg & 1.23& 19.52$\pm$0.27 & $ 97.14_{- 96.68}^{+ 58.47}$ & $52_{-27}^{+60}$ &$0.047^{+0.02}_{-0.012}$ & $140^{+110}_{-70}$ \\
\rxj & 0.06 & 20.73$\pm$0.11 & $ 4.87_{- 2.85}^{+ 1.80}$ & $2.4_{-1.4}^{+3.6}$& $0.007^{+0.005}_{-0.003}$ & $82^{+50}_{-30}$\\
\sdel & 0.04 & 21.97$\pm$0.19 & $ 1.94_{- 5.78}^{+ 1.45}$ &   $8.9_{-5.9}^{+17}$ & $0.03^{+0.03}_{-0.014}$ & $27^{+30}_{-17}$ \\
\sbs & 0.88 & 18.92$\pm$0.13 & $ 12.23_{- 7.15}^{+ 4.51}$ & -- & $0.17^{+0.05}_{-0.04}$ & $28^{+7}_{-6}$ \\
\einc & 0.9 & 17.90$\pm$0.44 & $ 9.71_{- 9.67}^{+ 4.85}$ & -- & $0.40^{+0.22}_{-0.14}$ & $17^{+4}_{-3}$ \\
\qone & 0.16 & 19.09$\pm$0.12 & $ 1.94_{- 1.93}^{+ 0.97}$ &  --  & $0.168^{+0.05}_{-0.04}$ & $10^{+10}_{-3}$\\
\qone\ \small
\citep{morgan12} & 0.16 & 19.09$\pm$0.12 & $ 10^{+20}_{-5}$ &  --  & $0.07^{+0.04}_{-0.02}$ & $50^{+30}_{-30}$\\
\end{tabular}
\end{table*}


For their sample of objects, \citet{morgan10} also provide magnitudes corrected
for strong lensing amplification (see also table \ref{tab:pro}) based upon
{\it HST} observations in the F814W filter. Flux calibration is based on the
paper of \citet{holtzman}, see section \ref{sec:mmdot}.


\subsection{Masses and emissivity slopes}

Several methods are used to estimate masses of supermassive black
holes. Most of them are model-dependent and suffer from
biases of different nature. For bright distant quasars, masses are usually
estimated either through photometrical data (bolometric luminosity is restored
from multi-wavelength observations) or by measuring the widths of broad emission
lines and the size of the emitting region by reverberation mapping
\citep{BmK}. While the first method relies heavily on {\it ad hoc }
assumptions about the mass accretion rate and accretion efficiency, the second
has a fundamental uncertainty connected to the geometry of the emitting
region. Virial mass is estimated as:

$$
M = f \frac{\sigma^2 R_{BLR}}{G},
$$

where $\sigma^2$ is the velocity dispersion corresponding to the observed line
width, $R_{BLR}$ is the size of the emitting region (determined with help of
reverberation mapping), and coefficient $f$
is calibrated using better-studied nearby active galaxies where $f\simeq
5.5$ \citep{onken}. { Sometimes only a limited number of spectra is
  available and reverberation analysis in impossible. In this case, empirical
  virial relations \citep{VP06} are used. These two types of mass estimates
  will be hereafter referred to as virial. }
Since we use the microlensing-based disc radii from \citet{morgan10} we also make
use of the virial masses given in this work (see references in this paper,
especially \citet{peng06}).

\section{Results}\label{sec:res}

\subsection{Masses and accretion rates}

\begin{figure*}
 \centering
\includegraphics[width=\textwidth]{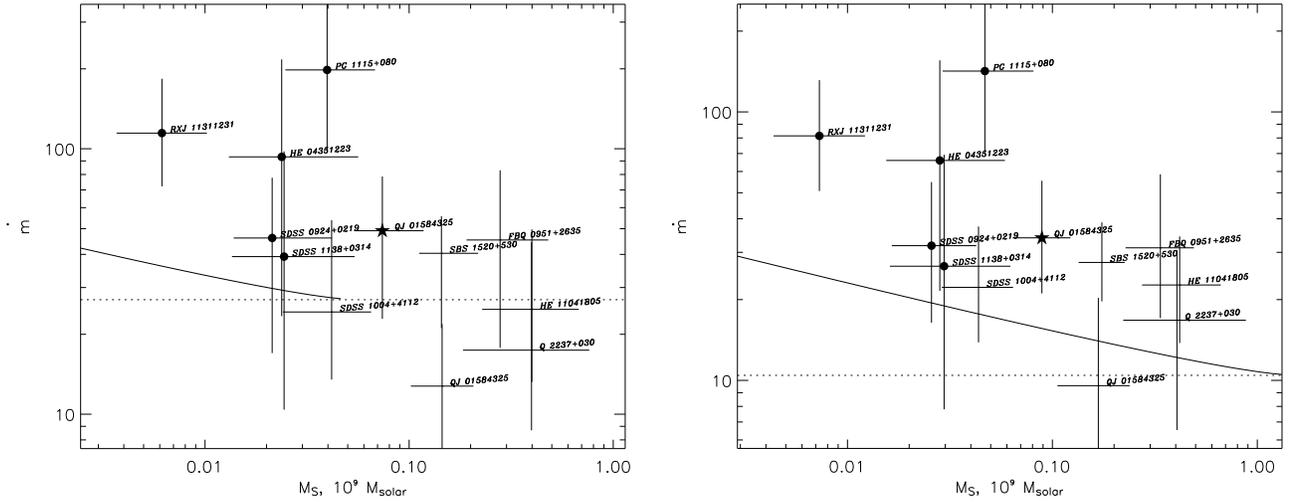}
\caption{ Quasar parameters (mass and dimensionless mass accretion rate) 
  recovered in the framework of supercritical
  envelope. The left and the right panels correspond to $a=0$ ($x_{in} =6$)
  and $a=0.9$ ($x_{in} \simeq 2.32$), respectively.  Horizontal dotted line
  shows the actual Eddington limit: accretion is sub-critical below the line
  and super-critical above it. Solid inclined line marks
  the limit where the envelope becomes larger than the disc monochromatic
  size (for $\lambda_{em} = 0.25\mu$). Hereafter, thick dots show objects
  that are also present in the sample of \citet{blackburne}, and the recent
  result for \qone\ \citep{morgan12} is shown by a star. {Errors
    correspond to  $1\sigma$ uncertainties in flux and radius.}
}
\label{fig:mmdot}
\end{figure*}

\begin{figure*}
 \centering
\includegraphics[width=\textwidth]{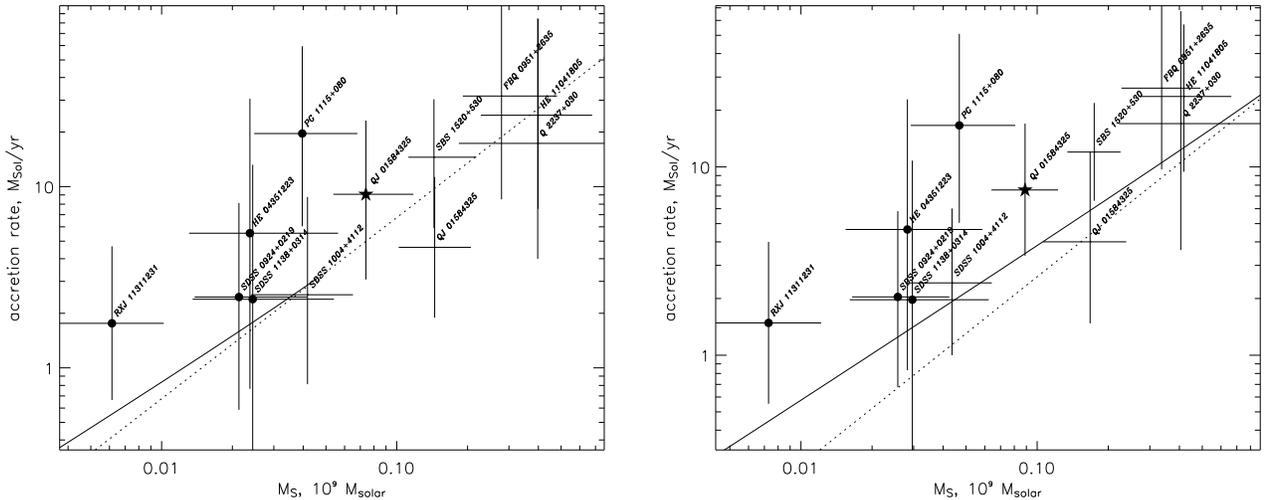}
\caption{Same as previous figure, but instead of dimensionless mass accretion
  rate \mdot, dimensional mass accretion rate $\dot{M}$ is given. 
}
\label{fig:mmdotphys}
\end{figure*}

Mass and dimensionless mass accretion rates estimated from the observables by
the method introduced in section~\ref{sec:mmdot} are given in
table \ref{tab:pro} (for $a=0.9$ and $x_{in}=2.32$) and shown in figures
\ref{fig:mmdot} and \ref{fig:mmdotphys}
for two values of $a$ and $x_{in}$. All the masses and mass accretion rates were
determined in the assumption of existence of an optically-thick scattering
envelope. They apply only to the objects where the disc is surrounded by an
envelope larger than the disc itself (i. e. above both the dotted and the
solid lines in figure \ref{fig:mmdot}). 

{
Errors given in table~\label{tab:pro} and figures~\ref{fig:mmdot} and
\ref{fig:mmdotphys} were calculated using direct non-linear error
propagation. We used the 1$\sigma$ uncertainties given by \citet{morgan10} for
the radii and fluxes. Since we do not know the exact probability distributions
of these quantities, it seems to be the most reasonable
approach. We substituted $R_1 \pm \Delta R_1$ and $I\pm \Delta I$ into
equations~(\ref{E:mdotiter}) and (\ref{E:fnumass}) and interpreted the highest and
the lowest values of \mdot\ and $M$ as the ends of some representative
confidence interval. 
}

{
Properties of the larger part of the objects shown in figure~\ref{fig:mmdot}
are consistent with accretion in a moderately super-Eddington regime. If the
accretion efficiency is high ($a\gtrsim 0.9$), all the objects may be
interpreted as super-Eddington. 
}
Most of the objects are however difficult to identify as super- or
sub-Eddington sources due to large uncertainties in \mdot. 
{
More probable super-Eddington objects such as \rxj\ and \pg\ tend also to
have lower $\zeta$ (see section~\ref{sec:zeta}). 
}
Einstein's cross, one
of the most probable sub-critical discs from the sample, is evidently among the
lowest-\mdot\ objects. According to our model, \qone\ should best conform to the
thin disc model. Indeed, disc size for this object reported in
\citet{morgan10} is in good agreement with
the theoretical predictions. However, the recent work of \citet{morgan12}
reports evidence for a larger disc size, several times larger
than (but still consistent at about $ 1.5\sigma$ confidence level) 
the standard model predictions based on the measured
amplification-corrected flux. Since dimensionless mass accretion rate is
proportional to the {envelope radius}, the larger radius 
makes the properties of \qone\ consistent with our supercritical disc model. 

\begin{figure}
 \centering
\includegraphics[width=\columnwidth]{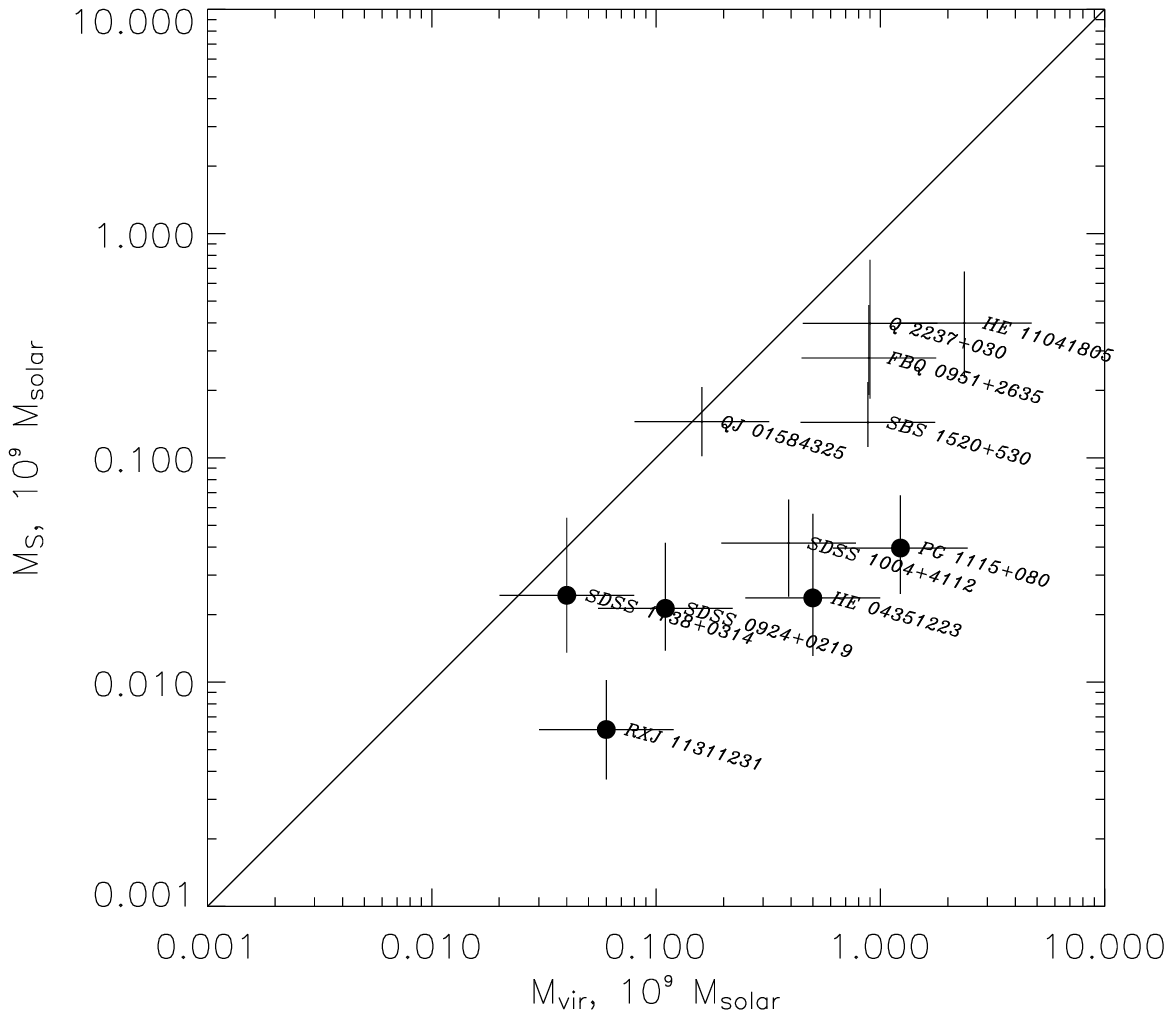}
\caption{ Virial masses { plotted versus mass  estimates in supercritical wind
  assumption }. Kerr parameter was set to $a=0.9$. }
\label{fig:mm}
\end{figure}

Masses determined in the spherical envelope model are generally smaller than
virial masses (see figure \ref{fig:mm}). Indeed, applicability of
the classical virial relations to an expanding supercritical wind is
questionable. As we will show in the section \ref{sec:balq}, the broad
emission lines observed in quasar spectra are unlikely to be formed in the
outflowing matter, but one
still may expect violations of the virial relations derived for sub-critical
active galactic nuclei. 


\subsection{Structure parameter correlation with mass}\label{sec:zeta}

\begin{figure}
 \centering
\includegraphics[width=\columnwidth]{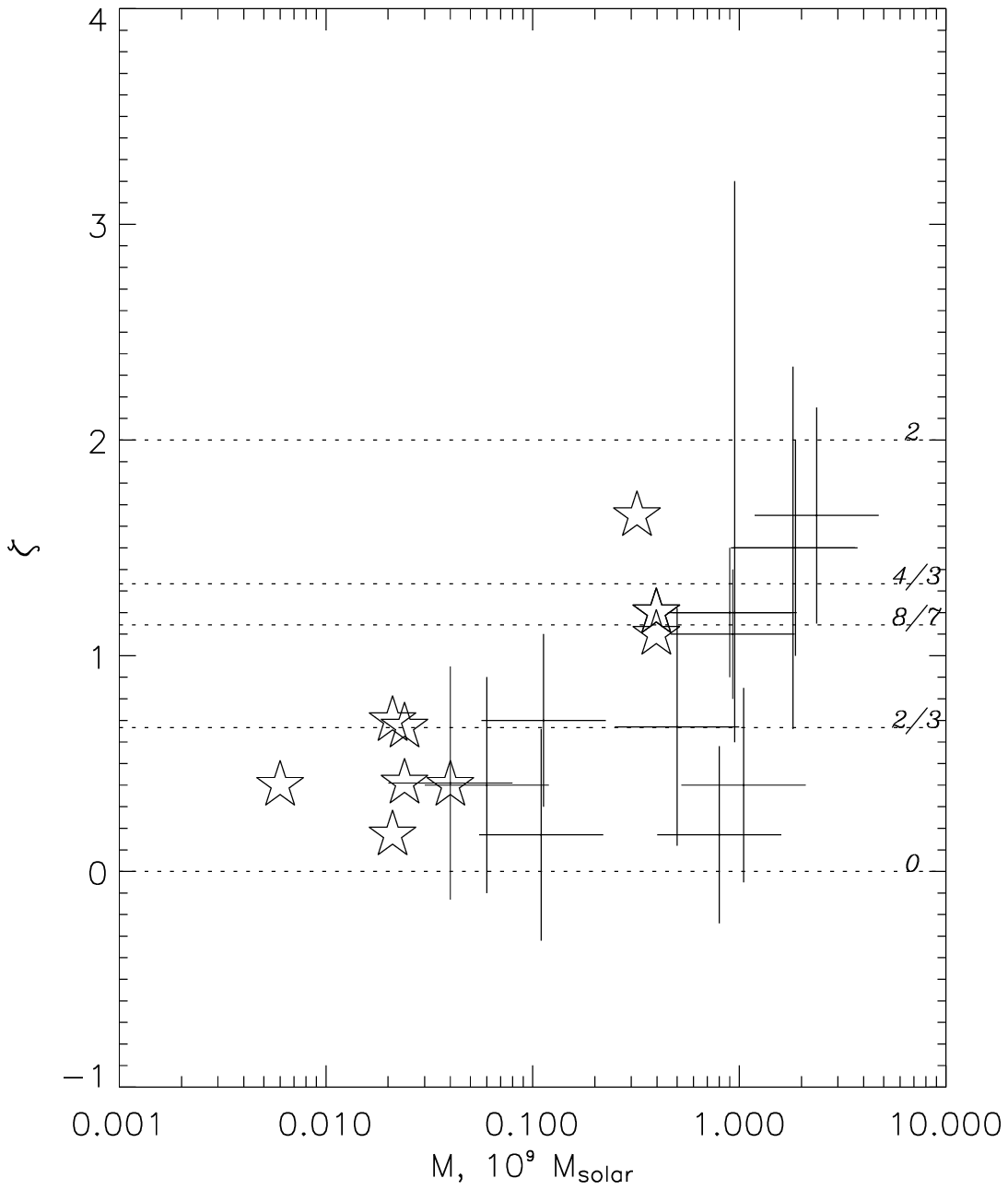}
\caption{ Structure parameters for quasars of different masses. Virial masses
  are shown by crosses. For several objects(\heone, \sdss, \heel, \pg, \rxj,
  \sdel, \einc), we use our envelope model to estimate masses (shown by stars). }
\label{fig:masszeta}
\end{figure}

Several quasar microlensing studies used multi-wavelength data to trace the
disc size dependence on wavelength. Fitting this dependence with a power law
allows to check the validity of several accretion disc models such as thin disc ($\zeta=4/3$),
slim or irradiated disc ($\zeta=2$) and a thin disc with a strong torque at the inner radius
($\zeta=8/7$). Available data are collected in table~\ref{tab:masszeta} and
in figure \ref{fig:masszeta}. 

Interestingly enough, \emph{all} the objects where standard disc slope works have
masses $M \gtrsim 10^9\Msun$. All the smaller-mass black holes show intensity
distributions with much lower $\zeta$. Besides the large fitting errors,
objects in figure \ref{fig:masszeta} may be separated in two groups: high-mass black
holes surrounded by accretion discs similar to standard and lower-mass objects
where $\zeta \sim 0\div 0.5$. An evident qualitative solution is
to propose that at least some quasars accrete in super-Eddington regime. If
all the bright lensed quasars accrete at $\dot{M} \sim 30\Msunyr$, Eddington
luminosity will be reached for $M\sim 10^9\Msun$ for accretion efficiency
$\eta \sim 0.1$. Lower-mass objects are
expected to enter super-Eddington accretion phase easier and 
lose excess accreting matter \citep{SS73}. We propose that supercritical wind
does not affect the spectral energy distributions of QSO much but changes its
spatial properties, acting as a lampshade that  changes only the visible size
and shape of the lamp. 

\begin{table}\centering

\caption{ Structure parameter as a function of SMBH mass. All the data were
  taken from \citet{blackburne}, if not stated otherwise. 
}
\label{tab:masszeta}

\footnotesize
\bigskip

\begin{tabular}{lccc}

Object & $M_{vir},$  & $\zeta$ & reference \\
 &   $10^9\Msun$ &  & \\
\hline
\mg & $1.82$ & $1.5\pm 0.8$ & \\
\mg & $1.82$ & $1.5\pm 0.5$ & \citet{bate08} \\
\heone & $0.50$ & $0.7\pm 0.6$ & \\
\rxn & $0.80$ & $0.17\pm 0.4$ & \\
\sdss & $0.11$ & $0.17\pm 0.5$ & \\
\sdss & $0.11$ & $0.7\pm 0.4$ &  \citet{floyd09} \\
\heel & $2.37$ & $1.65\pm 0.5$ & \citet{poindexter08} \\
\pg & $1.05$ & $0.4\pm 0.5$ & \\
\rxj & $0.06$ & $0.4\pm 0.5$ & \\
\sdel & $0.04$ & $0.4\pm 0.5$ & \\
\einc & $0.90$ & $1.2\pm 0.3$ & \citet{eigenbrodII} \\
\einc & $0.90$ & $1.1\pm 0.3$ & \citet{eigenbrodII}\\
&&& (no velocity prior) \\
\einc & $0.90$ & $1.2^{+2}_{-0.6}$ & \citet{anguita08} \\
\end{tabular}
\end{table}

\section{Discussion}\label{sec:balq}

In previous sections we have shown that some lensed quasars are
surrounded by scattering envelopes of moderate optical depths $\tau \sim R_1 /
R_{sph} \sim \mdot^{1/2} \sim 1\div 10$. Expected outflow velocities are $v
\sim \mdot^{-1/2} c$. At the same time, Doppler widths of broad emission lines
are significantly lower, $v/c \sim 0.01$. Besides, the expected emission line
luminosities are several orders of magnitude less than that of observed broad
emission lines in quasars. If we propose a constant filling factor $f$, a
recombination line in a supercritical outflow should have a luminosity
estimated as the following volume integral:

$$
L_{line} =\frac{1}{f} \alpha h\nu \times \int_V n_e n_i dV 
$$

Here, $\alpha$ is the recombination coefficient, 
$n_e$ and $n_i$ are the {electron concentration and concentration of
  the particular ion emitting the line}. It is convenient
to express the concentrations as $n_i = n_e x_i = \rho x_i / m$. Below, we fix
the values of $x_i$ and the effective particle mass $m$. { For
  completely ionised hydrogen-rich gas, $m$ is about the proton mass. } 

$$
\frac{L_{line}}{L_{Edd}} \simeq 10^{-7} \frac{f_w^2x_{i}\mdot^2 }{f\beta^2}
\frac{\alpha}{10^{-13} \rm cm^3 s^{-1}} \frac{1000\AAA}{\lambda_{line}}  \frac{R_{sph}}{R_{in}}
$$

Integration is performed from some inner radius $R_{in}$ to
infinity. Since the inner parts of the flow are considerably ionised the above
luminosity is an upper estimate. 
Predicted equivalent widths are of the order $\sim 10^{-4}\AAA$,
about five-six orders lower than the observed equivalent widths of broad
emission lines. Conditions are much more favourable for formation of
absorption lines, since the wind is thick to electron scattering and $N_H \sim
\tau / m_p \sim 10^{24} \mdot^{1/2} \rm cm^{-2}$. 

Outflows are expected to manifest themselves in blueshifted absorptions and P~Cyg
lines. Broad absorption line (BAL) quasars \citep{turnshek} show strongly blueshifted
absorption components of UV and sometimes X-ray spectral lines. 
Two of the objects of our sample, \pg\ and \sbs, are known as BAL quasars. 
\pg\ also demonstrates signatures of
moderately-relativistic outflows. In particular, \citet{chartas} find two
relativistic absorption components (at $\sim 0.1$ and $\sim 0.3 c$) of highly
ionised iron species and an OVI absorption component at $\sim 0.02c$. Similar
X-ray absorption systems were found for some other high-redshift luminous
quasars like APM~08279+5255 \citep{chartas02}, HE~J1414+117 \citep{chartas07}
and HS~1700+6416 \citep{lanzuisi}. 
Absorption components with relativistic velocities were also found in the UV
spectra of some BAL, ``mini-BAL'' and narrow-absorption line quasars (see
\citet{narayanan} and references therein). Relativistic outflows often
coexist with slower absorption systems, and UV and X-ray absorption lines
usually show different profiles and velocities. 

Discrepancy in wind velocities for different absorption components suggests
that the wind is highly inhomogeneous, with different components having
different velocities. Besides, its structure may be far from spherical
symmetry, and even in the spherically-symmetric case the shape of the
visible photosphere is distorted if the wind is relativistic
\citep{abram91}. For relativistic winds, there are two effects important for
their spatial properties: {\it (i)} firstly, relativistic beaming makes the
visible size of the emitting region $\sim \gamma^{3}$ times smaller, where
$\gamma$ is Lorentz-factor, and {\it (ii)} secondly, the optical depth along
the wind flow direction is smaller. Both effects are expected to produce a
wavelength-dependent limb-darkening effect that may be responsible for
deviations of $\zeta$ from 0 for some objects. 

For several objects, the size of the X-ray emitting region was studied using
microlensing effects (see \citet{pooley07,einc_xrays, chen12, morgan12} and
references therein). Independently of the UV/optical structure parameter
$\zeta$, the X-ray sizes of all the studied quasars are estimated as several
gravitational radii, that is sometimes one-two orders of magnitude smaller
than the proposed envelope size. 
This is difficult to account for in the spherical envelope model since the
size of the envelope {(as well as the size of the accretion disc in the
optical/UV range)} is generally much larger. 

However, the
outflows formed by super-Eddington accretion discs are expected to possess
high net angular momentum that leads to formation of an avoidance cone also
known as supercritical funnel \citep{SS73}. This picture is supported by
numerical simulations \citep{ohsuga, okuda}{: for a large range of
  inclinations, the angular size of the source in X-rays is considerably
  smaller than the size of the outer photosphere of the wind.} Optical depth of an envelope
with a funnel is expected to be much lower (if the disc is viewed
at low inclinations, see \citet{poutanen}) for the innermost parts of the
disc where the X-ray component is supposed to be formed. The non-monotonic optical
depth dependence on radius predicted for supercritical accretion disc winds
can qualitatively explain the decrease in the disc size at smaller
wavelengths observed for some objects with $\zeta \sim 0$ such as \pg\ and
\wf\ \citep{blackburne}.

{
As long as we use virial masses, properties of the X-ray radiation indicate
that it is formed near the last stable orbit \citep{morgan12}. However,
smaller masses recovered in the framework of our supercritical envelope model
shift the last stable orbit toward lower sizes. For instance, the X-ray
emitting region of \rxj\ has the size of $\sim 2\times 10^{14}\rm cm$
\citep{dai10}. For the virial mass estimate of $6\times 10^7\Msun$, this
corresponds to $\sim 7\times GM/c^2$, while the envelope-based mass is about
an order of magnitude larger, and the estimated X-ray size becomes several
tens of $GM/c^2$. This difference can hardly be used to distinguish between
the individual mass estimates or individual models of X-ray emission
separately. However, it should be borne in mind that virial masses are
consistent with the models where X-ray emission is formed near the last stable
orbit while the envelope-based mass estimates allow the X-ray emission to be
extended for tens of gravitational radii. This is consistent, for example,
with the models like \citet{liu12} where the X-rays are produced by the hot
gas of the corona present in the inner parts of the accretion flow. In any
case, the size of the X-ray emitting region is expected to be considerably
smaller than both the disc size at $\sim 2500\AAA$ and the scattering
wind photosphere size. 
}

Existence of true absorption processes should also affect the apparent size of
the spherical envelope. Assuming total thermalisation in a spherical
relativistic wind, \citet{fukue} find that the visible photosphere surface
follows the law $T(R) \propto R^{-1}$ that implies $\zeta \simeq 1$. More
elaborate studies taking into account the temperature and ionisation structure
of the wind are needed to explain the observed $\zeta \sim 0 \div 0.5$ values
of most of the putative supercritical quasars. 

True absorption processes are also important for wind acceleration. In the
super-Eddington regime, wind is efficiently launched by resonance lines even in
presence of strong X-ray radiation \citep{proga1,proga2}. Resonance lines do
not alter the measured photosphere size significantly since the wind is
opaque { to absorption } only in a narrow wavelength range. However,
their contribution to wind
acceleration through $\beta$ and $f_w$ may be important. 

It is tempting to compare the population of super-Eddington quasars to the
few known and well-studied supercritical black hole X-ray binaries, primarily
to SS433 \citep{ss433,cherep_integral}. For SS433, dimensionless mass
accretion rate is of the order several thousands that implies much slower
outflow velocity of $\sim 1000\kms$ and thermalised emission from
the wind pseudo-photosphere. 
On the other hand, mass accretion rates estimated in the present work, as well
as the values given by \citet{collin}, are considerably smaller. 
Maximal values are of the order $\mdot \sim
100\div 200$. Note that these values are only moderately supercritical,
Eddington luminosity is exceeded a factor of $\sim 10\div 20$ (depending on
the unknown accretion efficiency $\eta \sim 0.06\div 0.4$). It is more
instructive to compare the population of super-Eddington quasars to the
high-luminosity states of X-ray binaries like \grs\ \citep{grs_nedd} rather
than to persistent strongly supercritical accretors like SS433
or to sources like V4641 \citep{v4641} suffering strongly super-Eddington
outbursts. 


\section{Conclusions}\label{sec:conc}

Scattering envelope formed by a super-Eddington accretion disc
is a plausible model for the spatial properties of the emitting regions in
some lensed quasars. Large spatial sizes ($R\sim 10^{16}\div 10^{17}\rm cm$)
practically independent on wavelength are an expected outcome of a moderately
super-Eddington ($\mdot \sim 10\div 100$) mass accretion rate. Black hole
masses and mass accretion rates may be determined self-consistently if both
disc size and flux estimates are present. 

{
The small sizes of X-ray emitting regions of microlensed quasars may be
explained by existence of an avoidance cone, or supercritical
funnel, in the disc wind.
}

Some of our super-Eddington objects are broad absorption line
quasars, and at least one (\pg) shows signatures of a mildly relativistic
outflow. 

\section*{Acknowledgements}

We are grateful to R. A. Sunyaev for valuable discussions and for careful
attention to this work, and the Max Planck Institute for Astrophysics (MPA
Garching) for its hospitality.  Our work was supported by the
RFBR grant 12-02-00186-a. Special thanks to K. A. Postnov for useful comments
and discussions on BAL quasar outflows.

\appendix

\section{Brightness profile model}\label{sec:app}

\subsection{Radiation transfer in Eddington approximation}\label{sec:app:rt}

Optically-thick wind forms an extended pseudo-photosphere with a power-law
density slope, $\rho \propto R^{-2}$. In general, radiation transfer is
described by equation \citep{mihalas}:

$$
\mu \pardir{R}{I} + \frac{1-\mu^2}{R} \pardir{\mu}{I} = -\varkappa \rho(R) (S-I)
$$

Here, $I=I(R,\mu)$ is monochromatic intensity, $\mu=\cos \theta $ is the
cosine of the angle between the radius vector and radiation propagation
direction, $S=S(R)$ is source function. We consider only isotropic coherent
scattering that allows to equate source function to intensity averaged over
solid angle. We use moment approach, and use the first three radiation intensity
moments defined as:


\begin{equation}
J = \frac{1}{2} \int_{-1}^{+1} I d\mu
\end{equation}

\begin{equation}
H = \frac{1}{2} \int_{-1}^{+1} I \mu d\mu
\end{equation}

\begin{equation}
K = \frac{1}{2} \int_{-1}^{+1} I \mu^2 d\mu
\end{equation}

\begin{figure*}
 \centering
\includegraphics[width=\textwidth]{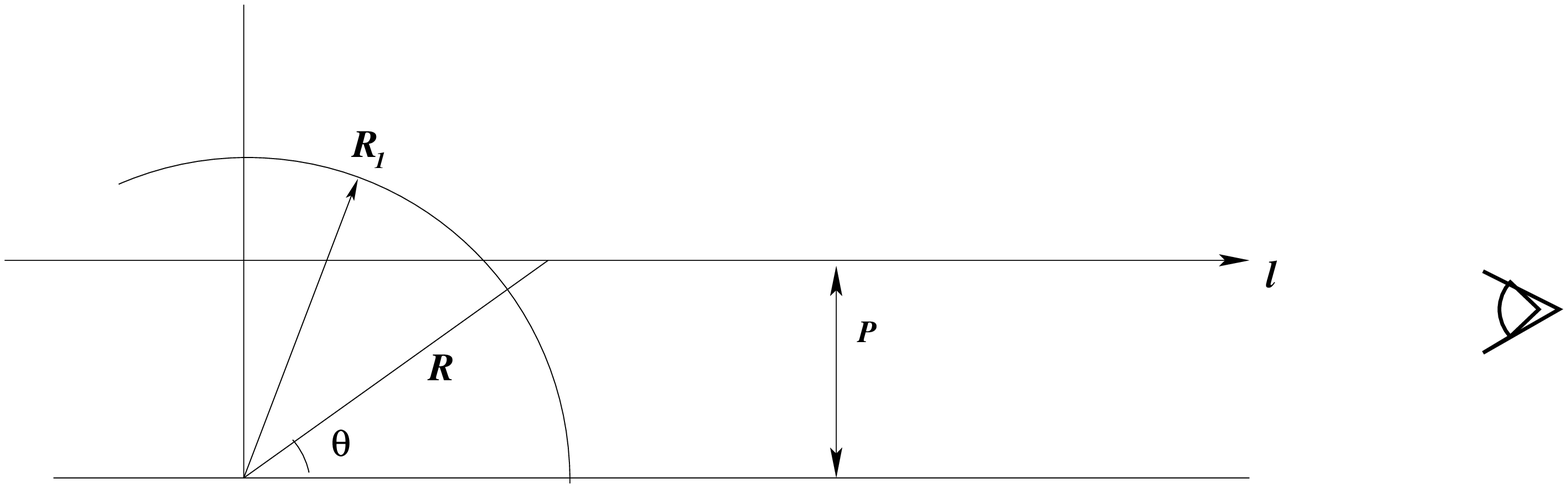}
\caption{Principal scheme illustrating integration along the line of sight
  performed in the Appendix.}
\label{fig:coordscheme}
\end{figure*}

First radiation moment $J$ has the physical meaning of mean intensity, hence
$S=J$ in our approximation. 
The system of moment equations is closed by Eddington's assumption $K=  fJ $
where $f=1/3$ valid in diffusion approximation. 
Questionability of Eddington
approximation for extended atmospheres is well known \citet{chapman} but it is
still
sufficient for our purposes. Here, we consider pure electron scattering by a
spherical atmosphere with electron density $n \propto \rho \propto R^{-2}$. 
In this case, the system of moment equations takes the form:

\begin{equation}
\left\{
\begin{array}{l}
\frac{1}{R^2} \der{R}{R^2 H} = 0\\
 \der{R}{f J} + \frac{3f-1}{R} J  = -\varkappa \rho H\\
\end{array}
\right.
\end{equation}

The system is simplified if $f=1/3$ (inner parts, $\tau\gg 1$) and if $f=1$
(opposite limit, $\tau \to 0$). Mean intensity (and hence source function)
scales in these two approximations as $\propto C_1 R^{-3} $ and $\propto C_2
R^{-2} \times (1+\tau)$, respectively. It is convenient to use the second
formula and to set $C_1 = C_2 = H_0 = H(\tau=1)$.
Both asymptotics are then naturally reproduced. $H_0$ may be connected to the
physical flux at the photosphere as $F(\tau=1)=4\pi H_0$ and luminosity as $L=(4\pi)^2 R_1^2
H_0$.  However, this approximate
formula does not conserve the total flux (intensity integrated over the solid
angle deviates from the total radiation flux calculated as $F(\tau=1)=4\pi H_0$)
that may result in systematic errors, hence we adopt the following form for
the source function:

$$
S(r) = H_0  r^{-2}\times  \left( 1+d\times r^{-1/2} + r^{-1}\right),
$$

where $r=R/R_1 = 1/\tau$, and $d$ is a free parameter.
Integrating source function for some
shooting parameter $P$ yields the observed intensity:

$$
I = \varkappa \int_{-\infty}^{+\infty} S\left(\sqrt{P^2+l^2}\right)
e^{-\tau(P,l)} \rho\left(\sqrt{P^2+l^2}\right) dl,
$$

where $\tau(P,l)$ is the optical depth along the current line of sight:

$$
\begin{array}{l}
\tau = \varkappa \int_{-\infty}^{l} \rho\left(\sqrt{P^2+l^2}\right) dl  =\\
= \frac{R_1}{P} \left(\atan \frac{l}{P} + \pi/2 \right)
\end{array}
$$

The coordinates and designations are shown in figure~\ref{fig:coordscheme}.

Finally, intensity distribution may be expressed as the following definite integral:

\begin{equation}\label{E:app:intty}
\begin{array}{l}
I(p) =  H_0 \times \left( u_{2}(p)+u_{3}(p)+d\times u_{5/2}(p)\right)
\end{array}
\end{equation}

where $x= l/ R_1$ and $p=P/R_1$, and:

$$
u_k(p) = p^{-(k+1)} \int_0^\pi e^{-\theta/p} \sin^k\theta d\theta
$$

Constant $d$ is tuned in a way to fit the integral flux value, $2\pi
\int_0^{+\infty} I(p) p dp = 4\pi H_0$. Numerical integration allows to
estimate the value of $d\simeq -0.097$. 

Half-light radius for this model may be estimated as $R_{1/2} = 1.063 R_1$ to
an accuracy of about $10^{-3}$. 
Setting $d=0$ is also a reasonable approximation:
it overestimates the flux by only about 5\%, while $R_{1/2}
\simeq 1.05 R_1$ in this assumption. 

\bibliographystyle{mn2e}
\bibliography{mybib}

\label{lastpage}

\end{document}